\documentclass[onecolumn, pra] {revtex4-1}    

\usepackage{graphicx,verbatim}
\usepackage{dcolumn}
\usepackage{bm}

\renewcommand{\v}[1]{\vec{\mathbf{#1}}}

\begin{document}


\title{Quantum-enhanced gyroscopy with rotating anisotropic Bose--Einstein condensates}

\author{L.M. Rico-Gutierrez}
\affiliation{ School of Physics and Astronomy, University of Leeds, Leeds LS2 9JT, United Kingdom }
\author{T.P. Spiller}
\affiliation{ Department of Physics, University of York, York YO10 5DD, United Kingdom }
\author{J.A. Dunningham}
\affiliation{ Department of Physics and Astronomy, University of Sussex, Brighton BN1 9QH, United Kingdom }


\begin{abstract}
High-precision gyroscopes are a key component of inertial navigation systems. By considering matter wave gyroscopes that make use of entanglement it should be possible to gain some advantages in terms of sensitivity, size, and resources used over unentangled optical systems. In this paper we consider the details of such a quantum-enhanced atom interferometry scheme based on atoms trapped in a carefully-chosen rotating trap. We consider all the steps: entanglement generation, phase imprinting, and read-out of the signal and show that quantum enhancement should be possible in principle. While the improvement in performance over equivalent unentangled schemes is small, our feasibility study opens the door to further developments and improvements.

\end{abstract}

\maketitle

\section{Introduction}
High-precision inertial navigation systems (INSs) are an important enabling technology \cite{Chatfield1997a}. The fact they do not rely on external references means that they are secure against deception or jamming, which is a critical weakness of global navigation satellite systems. Gyroscopes and linear accelerometers are the key components of INSs since the time integration of their signals allows for dead reckoning, i.e. the location and orientation of an object to be known at all times. There is therefore considerable interest in improving their precision and performance.

Current gyroscopes rely on mechanical rotating flywheels, vibrating microelectomechanical systems (MEMS), or light in ring lasers or fibre optic arrangements. The best available devices are based on ring lasers that make use of the Sagnac effect whereby light is sent in opposite directions around a closed loop and the two components acquire a relative phase dependent of the angular velocity of the loop \cite{Anderson1994a}. Such a technique can measure angular velocities with sensitivities of around $7.8$\, prad/s/$\sqrt{{\rm Hz}}$ \cite{Schreiber2013a}, however these rely on rings with impractically large areas. The Sagnac effect applies equally well to atoms but the relative sensitivity to phase of matter waves is much greater than in optical systems. So atomic ring gyroscopes offer the potential for improved precision as well as more practical ring sizes. This effect has been experimentally demonstrated \cite{Gustavson2000a} and, while this has not yet reached the sensitivity for light, future experiments should be able to surpass it \cite{Stockton2011a}.

Further improvements should be possible by making use of entangled states \cite{Hallwood2006a} in a quantum metrology scheme. It is known that entanglement can enable better measurement precision to be achieved with the same resources \cite{Giovannetti2006a, Giovannetti2004, Dowling2008, Pezze2009a, Holland1993a, Dunningham2002a, Gerry, Giovannetti2011, Higgins2007,Kacprowicz2010a,Oberthaler, Cooper2012a}. The trade off is that entangled states are difficult to create and manipulate. This means that any realistic scheme is likely to involve a sequence of small entangled states and hence the improvement is likely to be modest, though this may still prove to be important. Quantum metrology is also known to have other advantages such as when delicate samples are being measured \cite{Wolfgramm2013a, Aasi2013a}.

This paper demonstrates a Òproof-of-principleÓ of a quantum-enhanced atomic gyroscope. We consider the experimentally accessible system of an atomic Bose-Einstein condensate trapped in a rotating two-dimensional trap and investigate how the entangled state can be created, the rotation imprinted, and the final signal read-out, thus providing an interferometric protocol used to estimate external rotations. We are able to show that by making use of entanglement, an improved sensitivity to rotations should be able to be achieved in this system. While the improvement is modest, this study shows that quantum-enhanced atomic gyroscopes are a realistic prospect and paves the way to further investigations that will improve the advantage and address practical issues and constraints that will be important in delivering a new technology.

\section{Model}
The physical system we consider consists of a mesoscopic sample of $N$ bosonic atoms of mass $M$ in an axially symmetric harmonic potential, with frequency $\omega_\perp$ in the $xy$ plane and $\omega_z$ in the $z$ axis, interacting through hard-core-type elastic collisions. We take $\hbar \omega_z$ to be very large compared to the interaction energy so that the dynamics along the $z$ axis are frozen, i.e. all particles occupy the lowest axial energy level, thus rendering the gas effectively two-dimensional at sufficiently low temperatures. The trapped gas is rotated at angular frequency $\Omega$ around the $z$ axis with the aid of an external potential which in the rotating frame appears as an anisotropic quadratic potential in the $xy$ plane. The Hamiltonian in the rotating frame is given by 
\begin{widetext}
\begin{equation}
H=\sum_{i=1}^N\left(-\frac{\hbar^2}{2M}\nabla_i^2+\frac{1}{2}M\omega_\perp^2(x_i^2+y_i^2)+2A M\omega_\perp^2(x_i^2-y_i^2)-\Omega L_{zi}\right)+\frac{\hbar^2 g}{M}\sum_{j<i}^N \delta(\v{r}_j-\v{r}_i),
\label{theH}
\end{equation}
\end{widetext}
where $A(\ll 1)$ quantifies the degree of anisotropy, which is treated perturbatively in the calculations. Here, $L_{zi}$ is the angular momentum component in the $z$ direction of the $i$-th atom and the  $-\Omega L_z$ term transforms the Hamiltonian to the rotating frame, where $L_z=\sum_{i=1}^{N} L_{zi}$ is the total angular momentum of the condensate. Finally, $g$ is the dimensionless interaction coupling constant which quantifies the strength of two-particle interactions and is related to the 3D scattering length $a$ by \mbox{$g=\sqrt{8\pi} a / \lambda_z$}, where \mbox{$\lambda_z=\sqrt{\hbar/M\omega_z}$}. Hereafter, we use dimensionless variables with $\omega_\perp$ and $\sqrt{\hbar/(M\omega_\perp)}$ as units of frequency and length; consequently, energy and angular momentum are expressed in units of $\hbar\omega_\perp$ and $\hbar$, respectively.

In the isotropic case ($A=0$), the Hamiltonian is exactly diagonalisable in blocks of definite total angular momentum for which a suitable Fock basis is given by the eigenstates of the total angular momentum $|N_\mathbf{0}, N_\mathbf{1}, \ldots N_\mathbf{k}, \ldots \rangle$, where $N_\mathbf{k}$ specifies the number of atoms in the single-particle level labeled by the index $\mathbf{k}$ which represents a pair of quantum numbers $(n_\mathbf{k},m_\mathbf{k})$, the principal quantum number and the projection of the angular momentum respectively \cite{feder}.

When $A\neq 0$, the anisotropic term connects subspaces of given total angular momentum $L_z$ separated by two units of $\hbar$. If the anisotropy is small, a suitable basis to study the system is given by a truncated isotropic basis comprising the subspaces with $L\leq L_\mathrm{max}$, where the particular finite maximum angular momentum cut-off $L_\mathrm{max}$ is chosen to ensure the convergence for the energies and eigenstates of the Hamiltonian \cite{dagnino}. In addition, due to the absence of any parity-breaking terms in the Hamiltonian (i.e. ones that couple even and odd angular momenta) only subspaces with even angular momentum parity need to be considered.

Each subspace of definite total angular momentum can be further classified according to the concept of Landau levels. In the case of independent bosons, near the centrifugal limit $\Omega\approx\omega_\perp$, the energy levels are grouped in highly degenerate levels called Landau levels, roughly separated by an energy gap of $2\omega_\perp$. In this sense, any basis state is said to pertain to the $n_\mathrm{LL}$-th Landau level if $n_\mathrm{LL}=1+\sum_\mathbf{k} \left[n_\mathbf{k}+(|m_\mathbf{k}|-m_\mathbf{k})/2\right] N_\mathbf{k}$. The importance of this classification resides in the fact that for fast rotating and weakly interacting condensates, the atomic dynamics is restricted to the lowest Landau level (LLL), when the standard LLL validity criterion is satisfied, i.e. the mean interaction energy is much smaller than the energy separation $2\hbar\omega_\perp$ \cite{feder}. The LLL approximation corresponds to considering only basis states with particles exclusively occupying single-particle orbitals $n_\mathbf{k}=0$ and $m_\mathbf{k}\geq 0$. This greatly reduces the Hilbert space and allows for a more computationally tractable system. However, it has been recently shown that, although the LLL approximation can suffy for the purpose of studying the nucleation of vortices \cite{Fetter2009} or symmetry breaking \cite{dagnino}, it is necessary to go beyond this approximation in order to accurately describe the quantum states of the system that are useful for metrology \cite{ricoNJP}. In fact, the LLL approximation is completely unable to capture the possibility of creating $N00N$ states, whereas the inclusion of a larger basis allows for it, and reveals a much richer system amenable to quantum metrology. In the present work, a better approximation that goes beyond the LLL is considered by including more Landau levels in our calculations.   

\begin{figure}[t]
\includegraphics[trim=0cm 5cm 0cm 5cm, clip=true,width=0.4\textwidth]{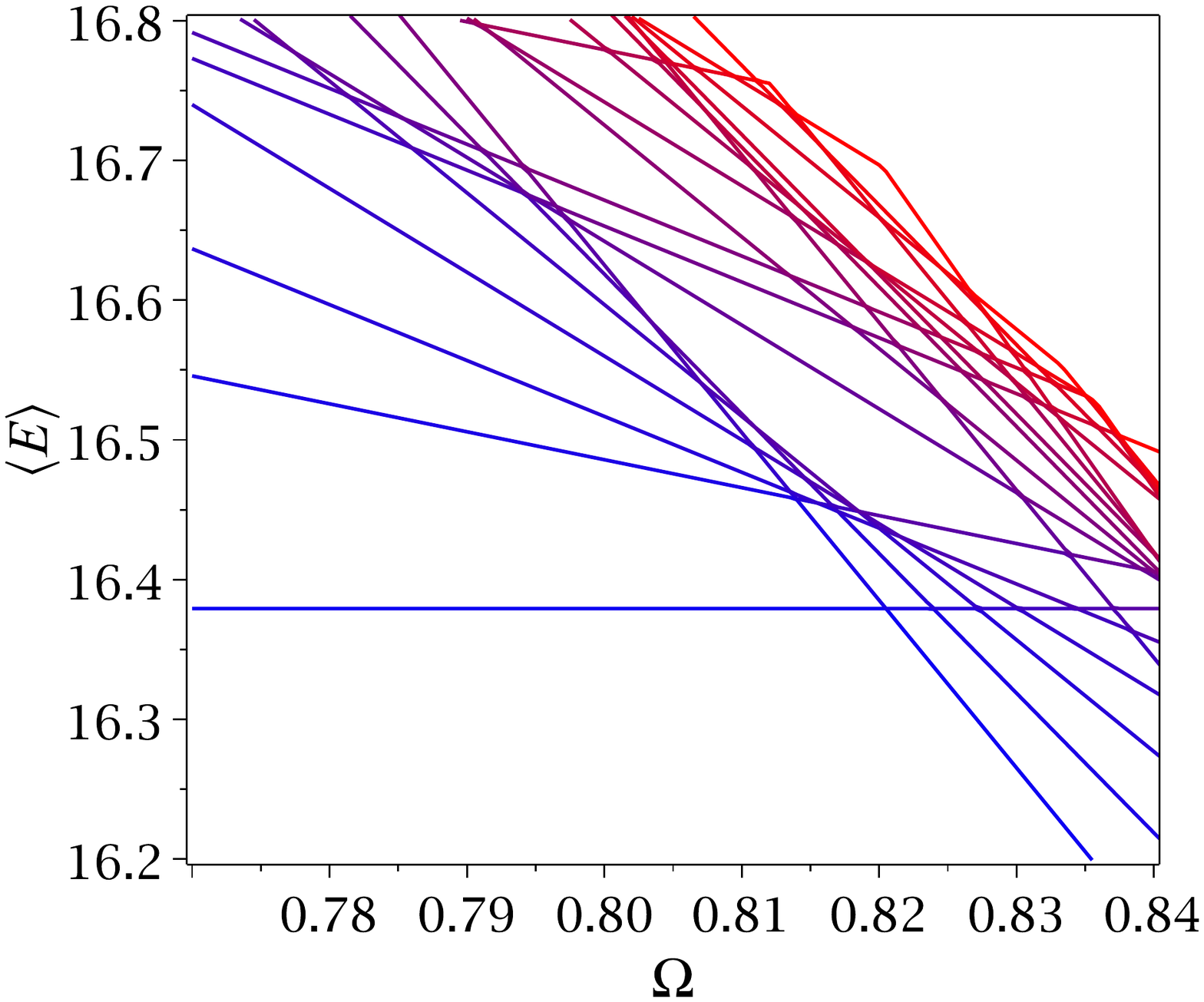}\hspace{0.05\textwidth}
\includegraphics[trim=0cm 5cm 0cm 5cm, clip=true,width=0.4\textwidth]{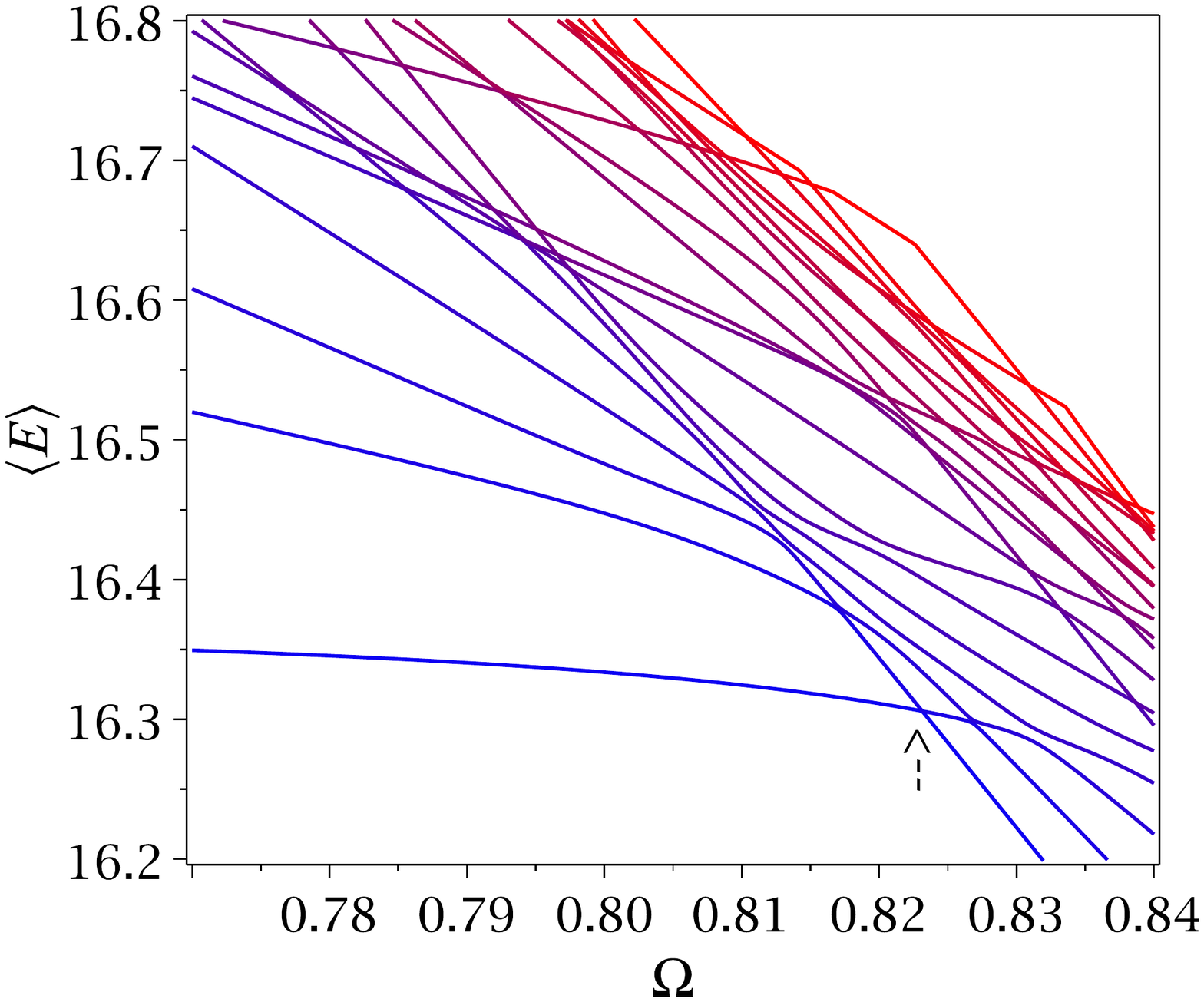}
\caption{\label{fig::energy} Energy spectrum as a function of the rotation frequency for $N = 12$
particles as calculated with a two Landau level approximation. (Left) Energy
spectrum for the axisymmetric case ($A = 0$). (Right) The energy levels for $A = 0.03$. In each case the interaction strength is given by $gN/6=1$ and $L_\mathrm{max} = N + 4$. An avoided crossing appears in the asymmetric case at the location indicated by the arrow.}
\end{figure}

Here, it is convenient to present the stationary energy spectrum of the system, as it provides a useful insight into the manipulation of the ground state considered to achieve the interferometric steps. The energy spectrum for $N=12$ particles obtained with a two-Landau-level basis and maximum angular momentum cut-off $L_\mathrm{max}=N+4$ is shown in Fig.~\ref{fig::energy}. Importantly, the inclusion of a non-vanishing anisotropic term in the Hamiltonian results in an avoided crossing between the ground state and the first excited state near a critical frequency for any number of atoms. The location of this is indicated by the arrow in Fig.~\ref{fig::energy}. As the condensate is adiabatically brought from being at rest to a rotation frequency just above the first avoided crossing $\Omega_\mathrm{min}$, where the ground state contains one vortex, the system passes a critical frequency $\Omega_c$ where the system undergoes turbulent symmetry breaking, heralding a quantum phase transition. Phase transitions are well known for showing/presenting an increase of correlations between particles, both in the classical context and the quantum one. Particularly, in the case of quantum phase transitions, critical phenomena is associated with scaling of entanglement in the vicinity of the transition point, which has been found to be a useful metrological resource that provides an improvement in the precision of the estimation of coupling constants and field strengths \cite{Fazio,Zanardi,ZanardiTwo}.

Exact diagonalisation results show that, for even numbers of particles, the ground state at $\Omega_c$ is a strongly correlated entangled state well described by a two-mode approximation
\begin{equation}
\label{twomodeapprox}
|\Psi_\mathrm{GS}(\Omega_c)\rangle=\sum_{n=0}^{N/2} C_n |N-2n\rangle|2n\rangle,
\end{equation}
where the two modes correspond to the most, and second most populated single-particle states $\psi_1$ and $\psi_2$, respectively. These two states are eigenstates of the single particle density matrix (SPDM) $\rho^{(1)}_{lk}=\langle \Psi_\mathrm{GS}|\hat{a}_\mathbf{k}^\dagger\hat{a}_\mathbf{l}|\Psi_\mathrm{GS}\rangle$ and their populations are equal at the critical frequency, which together account for most of the population of single-particle orbitals. Only the LLL single-particle states having $n_\mathbf{k}=0$ and $m_\mathbf{k}=0,1,2$ take part in the expansion of $\psi_1$ and $\psi_2$. Below and near $\Omega_c$, $\psi_1$ is a linear combination of the LLL states with $m_\mathbf{k}=0$ and $m_\mathbf{k}=2$, while $\psi_2$ is proportional to the state with $m_\mathbf{k}=1$. At $\Omega_c$, these two most populated states abruptly interchange their composition and remain the same up to a certain rotation frequency greater than $\Omega_c$. The two modes in Eq.(\ref{twomodeapprox}) for any ground state at criticality that can be obtained under variations of the different parameters considered in this work, can be shown to be entangled by using, for example, a Von Neumann entropy criterion for mode entanglement. More importantly, if the same state is expressed in the atom basis, it can also be shown that any one atom is entangled with the rest of them. Inter-atom entanglement is responsible for quantum enhancement in metrology protocols, even when mode entanglement is not present\cite{DemkowiczReview}.

The exact form of the two-mode approximation in Eq.(\ref{twomodeapprox}) significantly varies depending on the relative strength between the anisotropic perturbation and the interactions characterised by $g/A$ \cite{ricoNJP}. Exploration of the system for large values of $A$ requires a larger angular momentum cut-off, and thus a larger basis. In this work, we restrict the anisotropy strength to a fixed value of $A=0.03$ and varying values of $g\leq (6/N)$ so that the system can be well described with $n_\mathrm{LL}=2$ Landau levels and a maximum angular momentum $L_\mathrm{max}=N+4$. In principle, however, one can fix the interaction strength $g$ and vary the value of $A$ in order to get similar results for the exact ground state form.

\begin{figure}[t]
\includegraphics[trim=0cm 0cm 0cm 0cm, clip=true,width=0.4\textwidth]{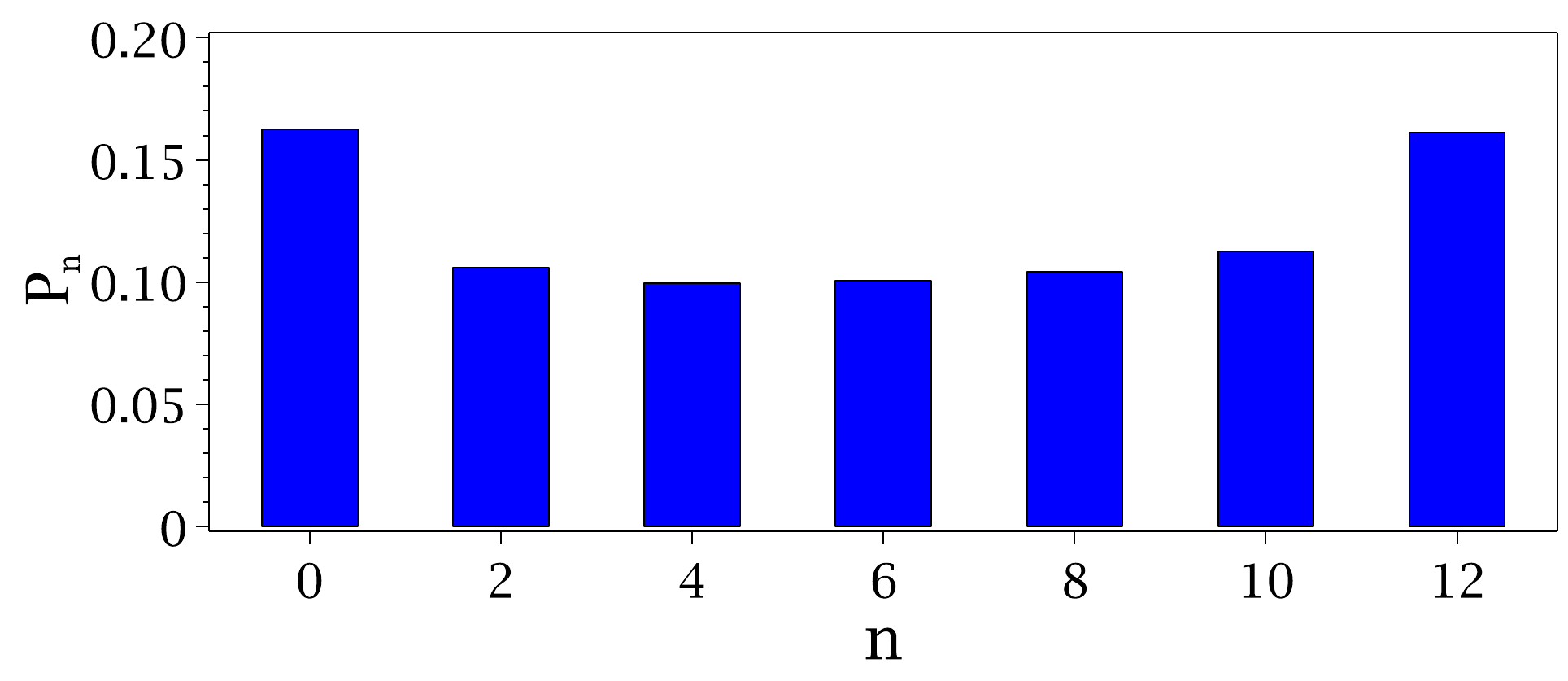}\hspace{0.05\textwidth}
\includegraphics[trim=0cm 0cm 0cm 0cm, clip=true,width=0.4\textwidth]{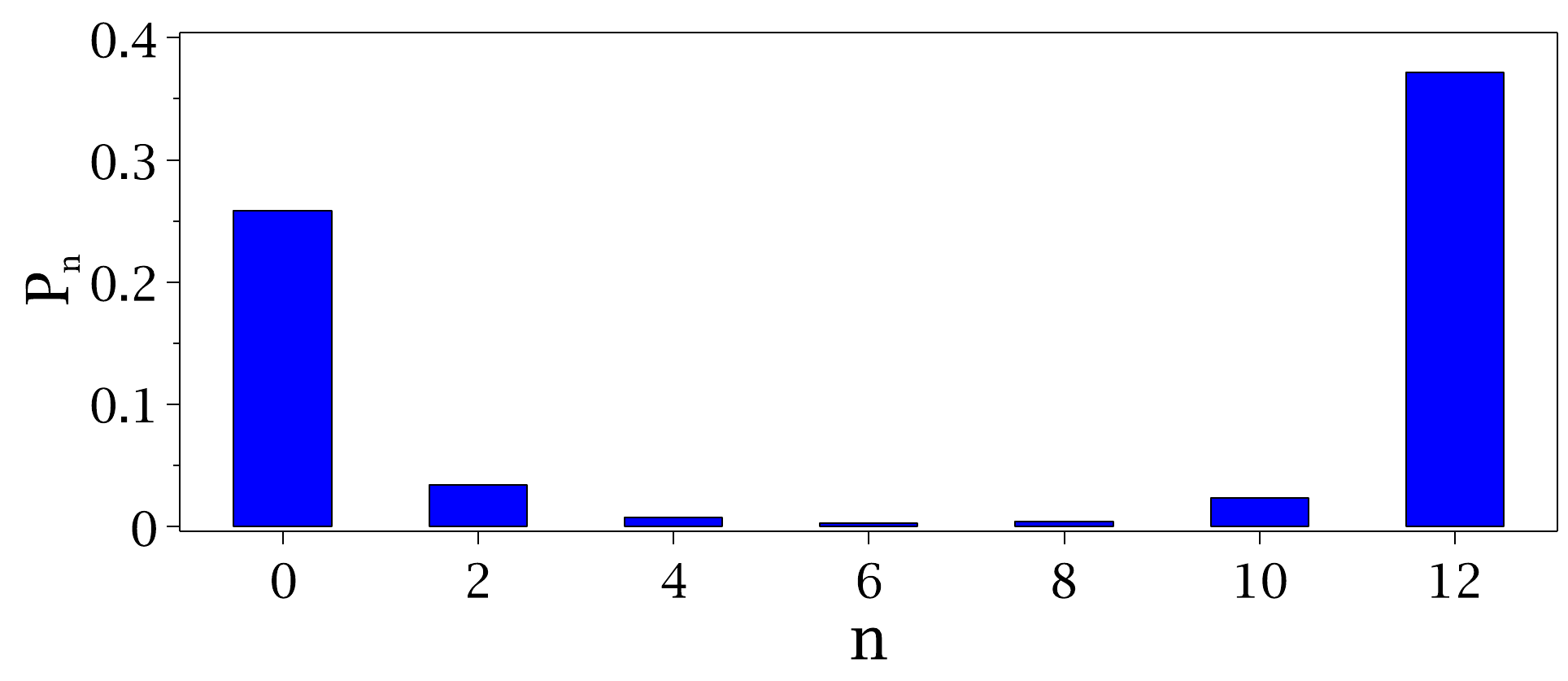}
\caption{\label{fig::batcat} The coefficients $P_n=\left|\langle N-2n,2n|\Psi_0(\Omega_c)\rangle\right|^2$ for the two-mode approximation at the critical frequency using two Landau levels for $N=12$ particles. (Left) For this bat-like state the interaction strength is given by $gN/6=0.44$, and the fidelity of the two-mode approximation is $\left|\langle \Psi_0(\Omega_c)|\Psi_\mathrm{TM}\rangle \right|^2=0.80$, where $|\Psi_\mathrm{TM}\rangle$ represents the two-mode approximation to the state. The calculated critical frequency is $\Omega_c\approx 0.938$. (Right) For the cat-like state, the interaction strength is $gN/6=1$, and the fidelity of the two-mode approximation is $\left|\langle \Psi_0(\Omega_c)|\Psi_\mathrm{TM}\rangle \right|^2=0.70$. The critical frequency in this case is $\Omega_c\approx0.823$. The strength of the anisotropy is $A=0.03$, and the truncation of the basis is $L_\mathrm{max}=N+4$ for both panels.}
\end{figure}

When the interaction strength $g$ is much smaller than $6/N$, the critical frequency $\Omega_c$ is closer to the harmonic trap frequency. In this case, in the absence of any anisotropy ($A=0$), the lowest-lying many-body eigenstates are energetically close together and the ground state at the first energy crossing is quasi-degenerate involving states with total angular momentum $0\leq L \leq N$, $L\neq 1$. The addition of the anisotropic perturbation  explicitly breaks the rotational symmetry and lifts the remaining degeneracy resulting in an entangled ground state at the critical frequency which is well approximated by a two-mode Holland--Burnett state, also known as a ``bat'' state, as shown in Fig.\ref{fig::batcat}; this ``bat'' state is the output state to a 50/50 beam-splitter having a twin-Fock state as an input \cite{Holland1993a}. The particular shape of this ground state is a direct consequence of the strong coupling of different quasi-degenerate states near the critical frequency connected by the anisotropic term, and the ``bat-like'' structure of this ground state is known to be robust to particle loss in interferometric schemes \cite{Cooper2010a}. 

Another important feature of this ``bat-like'' state is related to the concept of quantum Fisher information, which we briefly introduce next. The quantum Fisher information determines the lowest bound for the precision of any number of measurements of an undetermined parameter $\phi$ that has been encoded into the quantum state of a system, independent of the measurement scheme \cite{Caves}. The lowest bound is given by the Cr\'amer-Rao inequality
\begin{equation}
\Delta\phi \geq \frac{1}{\sqrt{n F_Q}},
\end{equation}
where $n$ denote the experimental repetitions, and $F_Q$ is the quantum Fisher information, which for a pure state is simply given by 
\begin{equation}
F_Q=4\left[\langle \Psi'(\phi) | \Psi'(\phi)\rangle - |\langle \Psi'(\phi)|\Psi(\phi)\rangle|^2\right],
\label{QFIpure}
\end{equation}             
with $|\Psi'(\phi)\rangle=\partial|\Psi(\phi)\rangle/\partial\phi$. Although the number of measurements $n$ is of key importance in reaching an optimal parameter estimation, in this paper we use the quantum Fisher information tool simply as a guide to inform the design of a practical scheme that shows a precision enhancement, which does not depend on the actual quantum Fisher information. Our parameter of interest is the rotation rate $\Omega$ and so we will consider the Fisher information as a function of $\Omega$. In the case of the ``bat-like'' state, the quantum Fisher information, after picking up an undetermined linear phase in the first mode, is peaked around the critical frequency and has a relatively broad bell-like shape \cite{ricoNJP}, which means that the unknown phase can in principle be determined with higher precision when encoded in the ground state of the system at any $\Omega$ contained in this relatively broad rotation frequency window. This is a convenient property since it means that an experimentalist would have a sizeable margin of error when trying to hit the critical frequency to prepare the initial entangled state for the interferometer.

On the other hand, for values of the interaction strength $g\approx 6/N$, the critical frequency is farther away from the harmonic trap frequency, and in the absence of any anisotropy, the first ground state energy crossing is a simple crossing between two many-body states having total angular momentum of $L=0$ and $L=N$. The larger value of the interaction strength has the effect of energetically separating the previously quasi-degenerate states so that only a simple crossing remains. In this case, when a non-zero anisotropy $A\leq 0.03$ is switched on, the perturbation term strongly mixes only the two mentioned states at the first ground state crossing. As a result, the degeneracy is lifted at the simple energy crossing, turning into an avoided crossing for which the ground state is well approximated by a two-mode ``cat-like'' state shown in Fig.~\ref{fig::batcat}. In contrast, the Fisher information as a function of $\Omega$ for the ground state in this case has a resonance-like shape, with a much larger value centred at the critical frequency, and has a width which is roughly two orders of magnitude smaller than the one for the ``bat'' state case. 

In short, for a fixed value of the anisotropy ($A\approx 0.03$), an interaction strength of $g\approx 6/N$ results in a cat ground state at the critical frequency $\Omega_c$, which exhibits a resonance-like shape for the quantum Fisher information as a function of $\Omega$. On the other hand, when the interaction strength has a value around $g\approx 3/N$, the resulting ground state at $\Omega_c$ is ``bat-like'' shaped, and the quantum Fisher information as a function of $\Omega$ has a broad bell-like shape centred around $\Omega_c$ \cite{ricoNJP}. This distinctive behavior of the system under changes in the interaction strength makes the rotating condensate a convenient system to create and tune a variety of different entangled ground states which can be used as initial states for quantum metrology schemes. In this article, we propose an interferometric scheme that uses the rotating condensate to measure external rotations, and we show that it can achieve sub-shot noise precision for all the different types of entangled states that can be produced by varying the interaction strength, which in turn can be achieved using Feshbach resonances \citep{Feshbach}. The interferometric scheme is conceptually simple and within reach of current technologies.

\section{Quantum interferometric scheme}

The interferometer is implemented as a quantum metrology protocol which can be divided into four main stages: adiabatic preparation of the initial entangled state, non-adiabatic coupling of the rotating condensate to the external system whose rotation frequency is to be measured, acquisition of an internal phase which encodes the external rotation into the ground state of the rotating condensate, and a read-out stage. Here, we show that after the phase acquisition, practical read-out schemes can determine the external rotation with a precision better than the classical (shot-noise) value for the parameters studied. 

The interferometric protocol requires the condensate rotation frequency to be varied with time. For simplicity, we assume a linear dependence with time as
\begin{equation}
\label{omegat}
\Omega(t)=\Omega_0+\gamma t,
\label{OmegaRamp}
\end{equation}
where $\gamma$ is the constant rate of change for the rotation frequency, and $\Omega_0$ is a particular initial rotation frequency. This linear dependence is the most simple and tractable way of modelling the dynamics of the system, however, it is flexible enough since any particular functional dependence of $\Omega(t)$ with time can, in principle, be approximated by a series of different linear passages. For the purpose of this work, we choose the simplest linear dependence in each stage in order to obtain a proof-of-principle result that exhibits sub-shot-noise precision, and thus, there is plenty of room for optimisation in this regard. In experiments, the variation of $\Omega(t)$ as it appears in Eq.(\ref{OmegaRamp}) corresponds to the variation of rotation frequency of the anisotropic potential ordinarly achieved by a stirring laser beam that rotates at the rate $\Omega(t)$ in the laboratory frame, and has a fixed anisotropic profile, $2A M\omega_\perp^2(x_i^2-y_i^2)$ in our case, in the rotating frame \citep{Gemelke,DalibardAngMom,GemelkeChu}.

In order to simulate the dynamics of the system, we follow \citep{dagnino} and solve the time-dependent Schr\"odinger equation $i\partial/\partial t|\Psi(t)\rangle=\hat{H}(t)|\Psi(t)\rangle$, where the only time dependency in $\hat{H}(t)$ comes from the term $-\Omega(t)\hat{L}_z$. Since in the absence of anisotropy ($A=0$), the Hamiltonian is exactly diagonalizable in blocks of definite total angular momentum with eigenvectors $|\Phi_j\rangle$ \cite{ricoNJP}, we expand $|\Psi(t)\rangle$ in the many-body angular momentum basis as $|\Psi(t)\rangle=\sum_i c_i(t)|\Phi_i\rangle$ and, by projecting the time-dependent Schr\"odinger equation onto the state $|\Phi_j\rangle$, we obtain a first order system of differential equations for $c_j(t)$,
\begin{equation}
i\frac{\partial}{\partial t} c_j(t) = \sum_i c_i(t) \langle\Phi_j| \hat{H}(t) |\Phi_i\rangle,
\end{equation}
subject to the initial condition $|\Psi(t=0)\rangle=\sum_i c_i(t=0)|\Phi_i\rangle$, which we solve numerically using a Fehlberg fourth-fifth order Runge-Kutta method. Due to computational constraints, all the numerical calculations in this article are restricted to $N\leq 12$ atoms, two Landau levels and a basis containing angular momentum subspaces of $L\leq N+4$.

The details of each stage in the interferometric protocol are discussed in the following sections. 

\subsection{Adiabatic preparation of the initial entangled state}    

\begin{figure}
\includegraphics[trim=0cm 5cm 0cm 5cm, clip=true,width=0.5\textwidth]{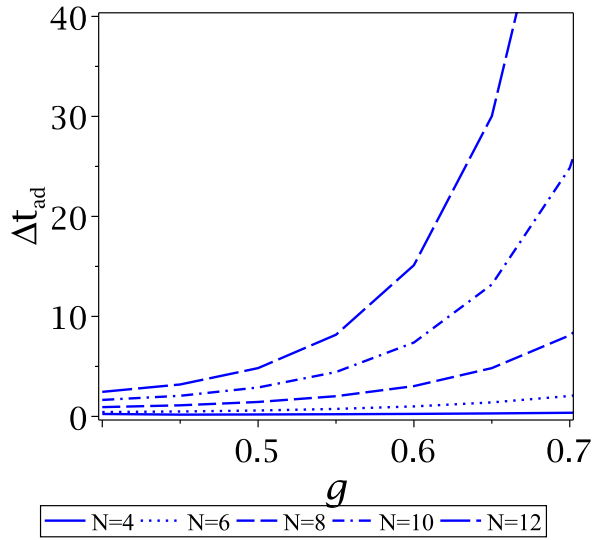}
\caption{\label{fig::AdiabTime} Predicted adiabatic evolution time as a function of the interaction strength for different numbers of particles.  
The full evolution from $\Omega=0.4$ to $\Omega=\Omega_c$ is divided in smaller linear segments with different angular acceleration rates as given by Eq.(\ref{gammaeq}), and a fixed increase in rotation frequency of 1\% of the trapping frequency. Results are shown for a trapping frequency of $\omega_\perp=2\pi \times 2.1$ kHz.}   
\end{figure}
  
The first stage of the protocol is the creation of the initial entangled state at the critical frequency $\Omega_c$. For this purpose, we start with a non-rotating condensate in the ground state, and adiabatically ramp up the rotation frequency until the condensate reaches the critical frequency. In experiments, the gas can be initially cooled down and condensed into the ground state at a moderate rotation rate $\Omega_0$ for which the energy gap to the first excited state is large enough to allow full condensation into the ground state by relaxation of the quantum gas \citep{dagnino}. Here, we assume this to be the case and consider nucleation of the entangled state by starting at $\Omega_0=0.4$ and adiabatically increasing the rotation frequency.

The main challenge of adiabatically nucleating the entangled state is to ramp up the rotation frequency from $\Omega_0$, slowly enough so that the system always remains in the ground state during the evolution, but fast enough so that the total evolution time is shorter than the average condensate lifetime. In order to estimate the minimum time needed to nucleate the entangled state, the full evolution from $\Omega_0$ to $\Omega_c$ is divided into $n$ smaller linear segments with fixed rotation frequency increments $\delta\Omega$, and corresponding rotation frequency change rates $\gamma_i$ which give the fastest evolution time for the $i$-th segment while preserving the adiabatic condition. For numerical calculations, we use $\delta\Omega=0.01$ and the maximum acceleration rates $\gamma_i$ are estimated using 
\begin{equation}
\label{gammaeq}
\gamma_i= \frac{\Delta E_\mathrm{min}^2}{N}\sqrt{p_{01}}, 
\end{equation}
where $\Delta E_\mathrm{min}$ is the smallest energy gap between the ground state and the next excited state for the $i$-th segment, and $p_{01}$ is the maximum tolerance for the transition probability between these two eigenstates after the adiabatic passage \citep{dagnino}, whose value we take to be $p_{01}=0.01$ for numerical calculations. In fact, Eq.(\ref{gammaeq}) is known to be quite restrictive as an estimate of the maximum acceleration rate consistent with the adiabatic condition \cite{messiah}, but it provides a relatively easy way to estimate adiabatic evolution times. 

We show the calculated minimum evolution time estimated using numerical results from exact diagonalisation for $\Delta E_\mathrm{min}$, and combining Eq.(\ref{omegat}) and Eq.(\ref{gammaeq}) in Fig.(\ref{fig::AdiabTime}) as a function of the interaction strength. A number of important conclusions can be drawn from these results. Firstly, nucleation of the entangled state via this adiabatic process is clearly limited to condensates containing small numbers of atoms, due to the fact that typical condensate lifetimes are of the order of $16$ s \cite{soding}. Thus, our scheme could be implemented in optical lattices of locally anisotropic rotating tight potentials ($\omega_\perp\approx 2\pi\times 2.1$ kHz) with thousands of sites with mean density between $N\approx 5$ and $N\approx 10$ atoms/site, which have been demonstrated in experiments \cite{gemelke}. Secondly, even for small numbers of atoms in relatively tight traps, the adiabatic evolution is feasible only for values of the interaction strength of $g<(6/N)\times 0.6$, which corresponds to the nucleation of bat-like entangled states. In contrast, practical adiabatic nucleation of cat-like states would require much tighter traps or great improvement in condensate lifetimes. Therefore, we mainly focus on the case of a bat-like state as the initial entangled state for our interferometer in the rest of the article. In actuality, this is not much of a limitation in quantum metrology schemes since it has been shown that bat-like states have the same potential of achieving nearly Heisenberg-limited precision as cat-like states in other geometries, even outperforming the latter when particle losses are considered \cite{Cooper2010a}. Moreover, for the particular geometry considered here, it has also been shown that a bat-like setup gives a more sizable margin of error in producing the entangled state, as opposed to the cat-like case \cite{ricoNJP}.

For notational purposes, we represent the operation performed during this first stage as
\begin{equation}
|\Psi_\mathrm{I}(\Omega=\Omega_c)\rangle=\hat{U}_\mathrm{A}|L=0\rangle,
\end{equation}
where $\hat{U}_\mathrm{A}$ represents the unitary operator corresponding to the adiabatic evolution, which transforms the non-rotating ground state $|L=0\rangle$ to the entangled ground state at the critical rotation frequency.

\subsection{Coupling to an external rotation}
The next stage is to expose the system to an external rotation that we wish to measure. This shifts the rotation frequency of the condensate in a rapid non-adiabatic way and takes it to a regime where its natural decomposition into eigenstates of the Hamiltonian at the shifted frequency allows the acquisition of an internal phase by means of simple free time evolution. This internal phase encodes information about the external rotation into the quantum state of the condensate, and the details of this acquisition are considered in the next stage of the protocol. 

The external rotation coupling can be modelled as having the following effect on the quantum state of the system 
\begin{equation}
\label{psiII}
|\Psi_\mathrm{II}(\Omega=\Omega_\Delta)\rangle=\hat{U}_\mathrm{ext}|\Psi_\mathrm{I}(\Omega_c)\rangle=\sum_i c_i(\Omega_\Delta)|\Psi_i(\Omega_\Delta)\rangle,
\end{equation}
where $\Omega_\mathrm{ext}$ is the external rotation we wish to measure, and $\Omega_\Delta=\Omega_c-\Omega_\mathrm{ext}$. Also, $|\Psi_i(\Omega_\Delta)\rangle$ is the $i$-th lowest lying eigenstate in the stationary energy spectrum of the Hamiltonian at $\Omega=\Omega_\Delta$, and $c_i(\Omega_\Delta)=\langle\Psi_i(\Omega_\Delta)|\Psi_\mathrm{I}(\Omega_c)\rangle$ are the expansion coefficients. We have made two important assumptions in Eq.(\ref{psiII}). The most important one is the assumption of very fast non-adiabatic evolution during the coupling to the external rotation, so as to obtain a high fidelity $\left|\langle \Psi_\mathrm{II}(\Omega_\Delta)|\Psi_\mathrm{I}(\Omega_c)\rangle\right|^2\approx 1$ between the initial entangled state and the final state after the coupling. This condition ensures that we retain the entanglement generated in the first stage before subjecting the system to a free time evolution. In fact, this is nothing but the main premise of enhanced quantum metrology, i.e., the encoding of unknown parameters that we wish to measure into an initial entangled state. In addition, this assumption provides a simple picture where the encoding of any information about the external rotation occurs only through the next stage in the interferometric protocol; that no information about the external rotation is gained during this sudden coupling is a consequence of $\partial|\Psi_\mathrm{II}(\Omega_\Delta)\rangle/\partial\Omega_\mathrm{ext}\approx 0$, resulting in a null Fisher information of the final state after the rotation frequency shift. Incidentally, this assumption also allows for shorter numerical simulation times. The second assumption consists in considering the external rotation to be small enough compared to the critical frequency, so that the resulting many-body state after the coupling can be described by a two-level superposition involving only the ground state and the next excited state at $\Omega_\Delta$ for a wide range of angular acceleration rates at which the coupling takes place. This is not a crucial assumption, since our numerical codes are not limited to this two-level approximation and our scheme does not rely on it. However, it will simplify the analysis in Section~\ref{sec:readout}.

 The sudden condition for the external rotation coupling imposes a limitation on the maximum value of the external rotation $|\Omega_\mathrm{ext}|$ that can be considered for a given maximum angular acceleration rate of the frequency shift. An estimate of the maximum value of the external rotation consistent with the assumption of non-adiabatic evolution can be obtained from the condition of the sudden approximation \cite{messiah} 
\begin{equation}
T\ll \frac{1}{\langle\Delta\overline{H}\rangle},
\end{equation}    
where $\langle\cdot\rangle$ denotes expectation values with respect to the initial state $|\Psi_\mathrm{I}(\Omega_c)\rangle$, and $T$ is the total time spent in the evolution, and $\langle\Delta\overline{H}\rangle^2=\langle\overline{H}^2\rangle-\langle\overline{H}\rangle^2$ is given in terms of the time-averaged Hamiltonian 
\begin{equation}
\overline{H}=\frac{1}{T}\int_0^T\hat{H}(t)dt.
\end{equation}
Combining these two expressions with Eq.(\ref{theH}), the sudden coupling assumption implies 
\begin{equation}
\label{suddencond}
|\Omega_\mathrm{ext}|\ll \sqrt{\frac{2\gamma_\mathrm{max}}{\langle\Delta\hat{L_z}\rangle}},
\end{equation}
where $\gamma_\mathrm{max}$ is the maximum angular acceleration rate of the frequency shift, and $\langle\Delta\hat{L_z}\rangle$ is the expectation value of the standard deviation of the total angular momentum with respect to $|\Psi_\mathrm{I}(\Omega_c)\rangle$. Since the initial entangled state (either bat-like or cat-like) consists of a superposition of angular momentum eigenstates with $L\leq N$, as our numerical calculations show, $\langle\Delta\hat{L_z}\rangle$ is at most of order $N$. Therefore, a larger number of particles further restricts the maximum value of the external rotation that is consistent with the sudden assumption. Nevertheless, since we have established that the first stage of the protocol already limits the feasibility of the scheme to small numbers of atoms, the $\sqrt{1/N}$ factor in Eq.(\ref{suddencond}) represents only a variation of less than one order of magnitude for $|\Omega_\mathrm{ext}|$. On the other hand, $\gamma_\mathrm{max}$ has a broader range of possible values. Here, we assume the external rotation to be less than one percent of the trap frequency in all calculations, for which Eq.(\ref{suddencond}) gives a maximum angular acceleration rate of the order of $\gamma_\mathrm{max}\approx 0.50\times 10^{-3}$. In fact, our numerical calculations show that a fidelity greater than $0.96$ can be achieved between the initial and final state for the external rotation coupling operation with $N\leq 12$ at this given maximum angular acceleration rate, and $|\Omega_\mathrm{ext}|<0.24\times 10^{-2}$.

\subsection{Phase imprinting}

The next stage of the interferometric protocol consists of a simple free time evolution at the shifted rotation frequency $\Omega_\Delta$, for a time $\tau$. The effect of this operation is that of creating a relative phase between the two components $|\Psi_0(\Omega_\Delta)\rangle$ and $|\Psi_1(\Omega_\Delta)\rangle$ that appear in the expansion of $|\Psi_\mathrm{II}(\Omega_\Delta)\rangle$ in Eq.(\ref{psiII}). This phase encodes information about the external rotation into the quantum state of the rotating condensate. 

The phase imprinting operation can be expressed as 
\begin{equation}
\label{psiIII}
|\Psi_\mathrm{III}(\Omega_\Delta; \tau)\rangle=\hat{U}_\mathrm{TE}(\Omega_\Delta;\tau)|\Psi_\mathrm{II}(\Omega_\Delta)\rangle=\sum_i c_i(\Omega_\Delta) e^{-i\Delta E_i(\Omega_\Delta)\tau}|\Psi_i(\Omega_\Delta)\rangle,
\end{equation}
where $\hat{U}_\mathrm{TE}(\Omega_\Delta;\tau)=\exp{(-i\hat{H}(\Omega_\Delta)\tau)}$ is the time evolution unitary operator, and $\Delta E_i(\Omega_\Delta)$ is the energy gap between the ground state and the $i$-th excited state at $\Omega_\Delta$. At this point, the value of the external rotation has been encoded into the quantum state of the condensate and it is ready to undergo a measurement stage.

\subsection{External rotation read-out} \label{sec:readout}
The read-out procedure is the final step and consists of undoing all the previous steps before the application of the phase. This is similar to the role of the second beam splitter in a standard Mach-Zehnder interferometer. After this, a suitable measurement is performed on the system to determine the external rotation. The particular choice of measurement scheme determines the precision with which the external rotation is estimated. Finding a practical way of optimising the measurement scheme is a rather hard problem, and there is no general strategy for it. However, a good starting point is to calculate the best precision possible using the quantum Fisher information, which provides a benchmark to aim for.

The fact that the quantum state of the system in Eq.(\ref{psiIII}) depends on $\Omega_\Delta$ in a rather non-trivial way, implies that there is no simple analytical expression for the quantum Fisher information of this state in general. However, a compact and insightful result can be obtained under some reasonable simplifying assumptions. Although our numerical simulations are not restricted to these assumptions, it is useful to have an analytical expression that provides additional insight and possibly serve as a guide to inform the design of the read-out stage. The first assumption consists of considering the waiting time $\tau$ to be small, so that we can neglect any terms of order $O(\tau^3)$ and higher. Secondly, we assume that at any point of the interferometric protocol, the population of excited states higher than the ground state and first excited state is negligible, thus making it possible to think of the system as a two-level one. This assumption is equivalent to considering only small external rotations, since the coupling to such small external rotations implies that, after the coupling, the quantum state of the system can be accurately reproduced by the apropriate superposition of the two lowest eigenstates of the Hamiltonian at the shifted rotational frequency. Under these assumptions, the quantum Fisher information right after the phase acquisition stage can be shown to be approximated by    
\begin{equation}
F_Q\left[|\Psi_\mathrm{III}(\Omega_\Delta;\tau)\rangle\right]\approx 4\tau^2 \langle\Psi_\mathrm{I}(\Omega_c)|(\Delta \hat{L})^2|\Psi_\mathrm{I}(\Omega_c)\rangle.
\label{FQapprox}
\end{equation}
This is a very interesting result. It implies that, as long as the waiting time and the external rotation are small enough, the lowest bound for the precision of rotation measurements is completely determined by the angular momentum fluctuations of the initial entangled state. A numerical calculation of the fluctuations of angular momentum shows that for the cat state, as well as for the bat state and intermediate states, $\langle(\Delta \hat{L})^2\rangle\propto N^2$, demonstrating that at least in this regime, the system is capable of delivering Heisenberg-limited scaling. As the waiting time starts to increase, going beyond the approximation used to obtain Eq.(\ref{FQapprox}), our numerical calculations suggest that the quantum Fisher information departs from the parabolic profile and shows oscillatory behaviour with respect to $\tau$, which is heavily dependent on the external rotation and particularly, on the number of atoms. Consequently, the scaling with $N$ beyond small waiting times becomes rather irregular. Therefore, finding the optimal waiting time involves calculating the longest waiting time that provides a better signal, but does not cross over to the region of irregular scaling with $N$.  An analysis of the optimal measurement is beyond the scope of the present work.

This result for the quantum Fisher information suggests that we look for a measurement of total angular momentum in a basis where the final output of the interferometer has a large variance in angular momentum in order to get close to the lower bound given by Eq.(\ref{FQapprox}).

One possible way to accomplish this measurement scheme is the following. After waiting for a time $\tau$, the condensate is quickly decoupled from the external rotation, meaning that the condensate returns to the rotation rate $\Omega_c$ under the sudden approximation. At this point, the two lowest-lying energy levels have the largest population. Then, we adiabatically switch off the anisotropy, thus bringing the system to a basis where the eigenstates of the Hamiltonian are also eigenstates of the total angular momentum with the ground state having $L=N$ and the first excited state $L=N-k$, where the value of $k$ depends on the magnitude of the interaction strength. For $gN/6\ge 0.6$ the value of $k=N$, and for smaller values of $g$, we have $k=1$. After the adiabatic anisotropy switch-off, we end up with a condensate occupying mostly the first two lowest-lying energy levels, which in the case of $gN/6\ge 0.6$ means that this final output state right before an angular momentum measurement has a large variance in $L$.

We now proceed to calculate the external rotation precision obtained from a total angular momentum measurement after bringing the system to the measurement basis mentioned above. In this case, it suffices to consider projective measurements corresponding to the projective measurement operators $\hat{E}(L)=|\tilde{\Psi}_i\rangle\langle\tilde{\Psi}_i|$, where $|\tilde{\Psi}_i\rangle$ is the $i$-th energy eigenstate at $\Omega_c$ when $A=0$, having $\hat{L}|\tilde{\Psi}_i\rangle=L|\tilde{\Psi}_i\rangle$, and thus $L$ labels the results of the measurement, in this case, a well-defined total angular momentum. The probability density for result $L$, given the external rotation $\Omega_\mathrm{ext}$, is
\begin{equation}
p(L|\Omega_\mathrm{ext})=\langle \Psi_\mathrm{IV}(\Omega_c; \tau)|\hat{E}(L)|\Psi_\mathrm{IV}(\Omega_c; \tau)\rangle,
\label{ProbDens}
\end{equation}
where $|\Psi_\mathrm{IV}(\Omega_c; \tau)\rangle$ is the final output state to the interferometer protocol right after the adiabatic anisotropy switch-off.

Consider now $n$ measurements of total angular momentum with results $L_1,\ldots, L_n$ which are randomly distributed according to the distribution in Eq.(\ref{ProbDens}). The external rotation is estimated using the function $\L_\mathrm{est}=\sum_{i=1}^n L_i /n$, which is nothing but the mean value of the total angular momentum of the sample. This function is called the estimator function, and how precisely the $n$ measurements are able to determine the external rotation depends on its particular functional form. Before the external rotation can be estimated using this function, it needs to be ``corrected'' to account for the difference in units \cite{Caves}. Thus, the external rotation estimation given by the units-corrected mean value of the sample distribution of the estimator is
\begin{equation}
\langle\Omega_\mathrm{est}\rangle=\frac{\langle L_\mathrm{est} \rangle}{\left|d \langle L_\mathrm{est}\rangle / d\Omega_\mathrm{ext}\right|},
\end{equation}
where $\langle L_\mathrm{est} \rangle$ is the mean value of the sample distribution of the estimator function $\L_\mathrm{est}$. It is worth emphasising that this is not a quantum mechanical expectation value, but a statistical one. Nevertheless, due to the fact that the estimator functional form is an arithmetic mean, we can invoke the central limit theorem for large samples with $n\gg 1$, in this case 
\begin{equation}
\langle L_\mathrm{est} \rangle=\langle\hat{L}\rangle,
\end{equation}   
where the right-hand side average is a quantum expectation value giving the mean value of the ``population'' distribution assumed within the central limit theorem context. In the same way, the variance of the sample distribution can be taken as
\begin{equation}
\langle (\Delta L_\mathrm{est})^2 \rangle=\langle (\Delta \hat{L})^2 \rangle / n,
\end{equation}
according to the central limit theorem. Consequently, the precision with which the external rotation can be estimated with this measurement scheme is given by the error-propagation formula
\begin{equation}
\Delta\Omega_\mathrm{ext}=\sqrt{\langle(\Delta \Omega_\mathrm{ext})^2\rangle}=\frac{\sqrt{\langle (\Delta \hat{L})^2 \rangle}}{\sqrt{n}\left|d \langle L_\mathrm{est}\rangle / d\Omega_\mathrm{ext}\right|}.
\label{DeltaOmega}
\end{equation} 

\begin{figure}
\includegraphics[trim=0cm 5cm 0cm 5cm, width=0.5\textwidth]{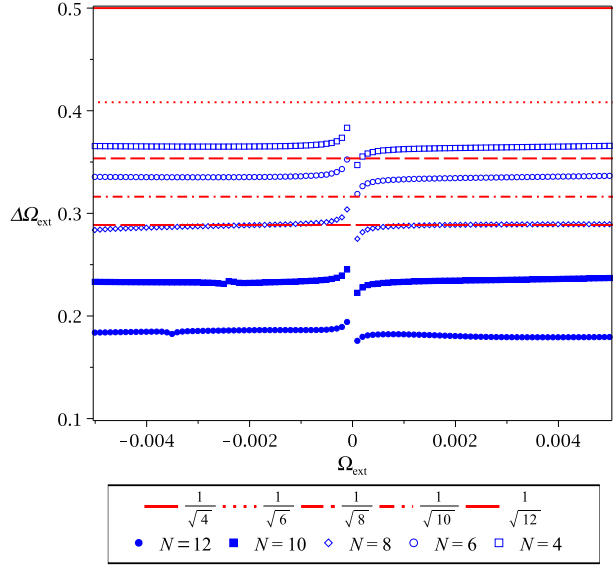}
\caption{\label{fig::DeltaOmtau10} Standard deviation for the distribution determining the external rotation as a function of the same rotation. Here, the estimator used to build the distribution is the total angular momentum of the interferometer output. In addition, the standard deviation has been scaled by a factor equal to the waiting time, in this case $\tau=10$, in order to benchmark against the shot-noise limit for unentangled particles. Also, $gN/6=0.44$ and $A=0.03$.}
\end{figure}

We show the precision obtained from a full simulation of the interferometric protocol for a particular set of parameters in Fig.(\ref{fig::DeltaOmtau10}), where sub-shot noise precision is achieved for the measurement scheme discussed above. In Fig.(\ref{fig::DeltaOmtau10}), we have factored out $\sqrt{n}$ appearing in Eq.\ref{DeltaOmega}, as well as $\tau^{-1}$, which are both independent of the number of particles, in order to directly compare the obtained precision with the shot-noise limit $1/\sqrt{N}$. This means that the precision shown in the mentioned figure above can be improved by a factor corresponding to the evolution time $\tau$ and $1/\sqrt{n}$, if the protocol is repeated $n$ times. Here, it is worth mentioning that any given estimator generally performs differently for different values of the unknown parameter, and this can be seen in our calculation corresponding to Fig.(\ref{fig::DeltaOmtau10}). In particular, the divergence at $\Omega_c$ is a direct consequence of a finite numerator and a denominator that goes to zero in Eq.(\ref{DeltaOmega}). The error-propagation formula is known to result in divergences like this for particular values of the unknown parameter. However, these divergences are not fundamental and can be avoided in different ways, for instance, by implementing a deliberate bias in the unknown parameter or by using a different estimation strategy, such as Bayesian estimation. Although we have only presented results for a frequency window about $\pm 0.5\%$ of $\omega_\perp$ around the critical frequency, our simulations show that sub-shot-noise scaling remains up to external rotation values in a frequency window $\pm 5\%$ of $\omega_\perp$ wide. However, meeting the criterion for the sudden approximation for larger external rotations implies a larger rotational frequency acceleration, which might not be realistic. Additionally, we have also observed similar sub-shot-noise scaling for larger values of the interaction strength $1\geq gN>0.44$.    

\begin{figure}
\includegraphics[trim=0cm 5cm 0cm 5cm, width=0.5\textwidth]{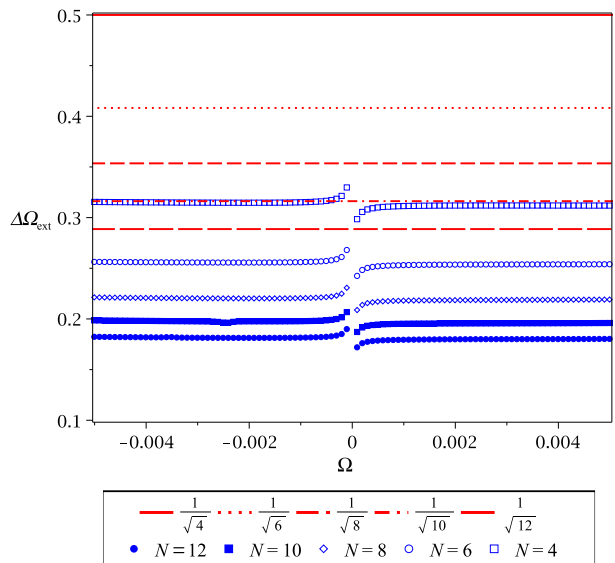}
\caption{\label{fig::DeltaOmtau10gs} Standard deviation for the distribution determining the external rotation as a function of the same rotation. Here, the estimator used to build the distribution is the population of the ground state of the interferometer output. In addition, the standard deviation has been scaled by a factor equal to the waiting time, in this case $\tau=10$, in order to benchmark against the shot-noise limit for unentangled particles. Also, $gN/6=0.44$ and $A=0.03$.}
\end{figure}

\begin{figure}
\includegraphics[trim=0cm 5cm 0cm 5cm, width=0.5\textwidth]{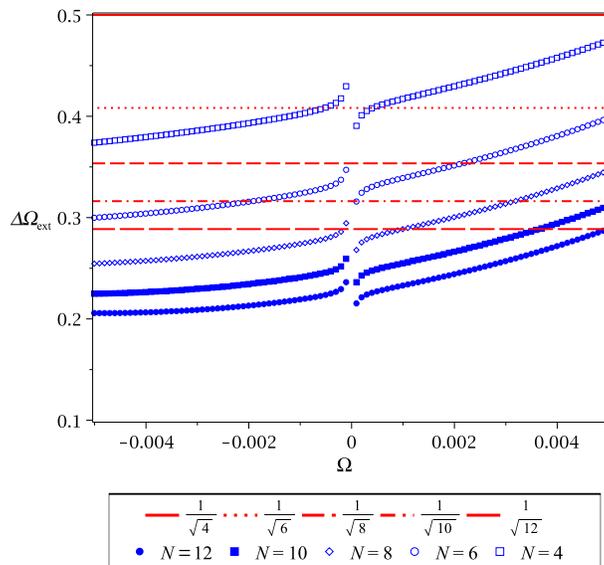}
\caption{\label{fig::DeltaOmtau50gs} Standard deviation for the distribution determining the external rotation as a function of the same rotation. Here, the estimator used to build the distribution is the population of the ground state of the interferometer output. In addition, the standard deviation has been scaled by a factor equal to the waiting time, in this case $\tau=50$, in order to benchmark against the shot-noise limit for unentangled particles. Also, $gN/6=0.44$ and $A=0.03$.}
\end{figure}

Although the measurement strategy mentioned above can deliver sub-shot noise precision, it presents a serious challenge for experimentalists, due to the adiabatic anisotropy switch-off. In contrast to the adiabatic nucleation of the entangled state, we need to preserve the population of several low-lying energy eigenstates, since the state of the condensate right after the decoupling from the external rotation can contain non-negligible overlaps with higher excited states, particularly for larger values of the waiting time $\tau$. Since very short waiting times are not suitable for practical reasons, it is advantageous to find an alternative measurement scheme. An obvious choice is a binomial estimation, where the two outcomes correspond to the probability of finding the system in the ground state or in any other excited state, regardless of which one exactly. In this case, the approach is exactly the same as before, but this time, we have only two measurement operators
\begin{equation}
\hat{E}_0(x=0)=|\tilde{\Psi}_0\rangle\langle\tilde{\Psi}_0|,
\end{equation}
and 
\begin{equation}
\hat{E}_1(x=1)=1-|\tilde{\Psi}_0\rangle\langle\tilde{\Psi}_0|,
\end{equation}
where $x$ is a random variable representing the outcome of a measurement in this second scheme, which can be $x=0$ if the condensate is found in the ground state, or $x=1$ otherwise. Taking the estimator function as an arithemetic mean as before, $X_\mathrm{est}=\sum_{i=1}^n x_i$, we can again invoke the central limit theorem and calculate the units-corrected standard deviation of the sample distribution for $\Omega_\mathrm{est}$ corresponding to this second approach. Notably, the precision obtained with this approach is slightly better than the first one but, more importantly, is more amenable to experiments. Full simulation results of the interferometric protocol using this second approach are shown in Fig.(\ref{fig::DeltaOmtau10gs}) for the same set of parameters as in Fig.(\ref{fig::DeltaOmtau10}). Although we have not pursued a detailed study of the dependence of the interferometric protocol on the waiting time, it is worth mentioning that increasing waiting times can have a significant impact on the precision obtained from the protocol, as it is illustrated by the particular example of Fig.(\ref{fig::DeltaOmtau50gs}).

\section{Conclusions}
In this paper, we have presented a proposal for an interferometric protocol to measure external rotations using a rotating anisotropic Bose-Einstein condensate. For simplicity we have assumed that particle losses are rare and can be neglected. Future work will consider the effect of losses, which connect the odd and even number subspaces and so complicate the dynamics. The protocol we have presented should be able to deliver sub-shot noise measurements as demonstrated by the quantum Fisher information of the system after an external rotation has been applied to it. Importantly, the feasibility of sub-shot noise measurements using two different measurement schemes was demonstrated for a particular set-up, which is a real strength of the presented protocol. Out of these two measurement schemes, the one that delivers the best precision is a simple projective measurement which only requires measuring the population of the ground state of the system at the output to the interferometer. Although it has not been presented in this study, a Bayesian estimation strategy can also be used to deliver the same results presented in this paper. The Bayesian strategy has the advantage of allowing any prior knowledge of the unknown external rotation to be accounted for, such as any previous measurement of the external rotation. Additionally, the interferometer is highly tunable, including the form of the entangled state generated inside of it, and it is amenable to a proof-of-principle experiment. Finally, in this investigation, we have barely scratched the surface of what the system could be capable of doing, as a full exploration of all its parameters is still under current investigation.    

\begin{acknowledgments}
We would like to acknowledge discussions with J. Cooling. This work was supported by the Defence Science and Technology Laboratory (UK), Consejo Nacional de Ciencia y Tecnolog\'ia (CONACyT, Mexico) and Secretar\'ia de Educaci\'on P\'ublica (SEP, Mexico).

\end{acknowledgments}




\begin{thebibliography}{20}

\bibitem{Chatfield1997a} A. Chatfield, {\textit{Fundamentals of High Accuracy Inertial Navigation}} (AIAA, Reston, 1997). 
\bibitem{Anderson1994a} R. Anderson, H.R. Bilger, G.E. Stedman, Am. J. Phys. {\bf 62}, 11 (1994).
\bibitem{Schreiber2013a} K. U. Schreiber and J.-P. R. Wells, Rev. Sci. Instr. {\bf 84}, 041101 (2013). 
\bibitem{Gustavson2000a} T. L. Gustavson, A. Landragin and M. A. Kasevich, Class. Quantum Grav. {\bf 17}, 2385 (2000).
\bibitem{Stockton2011a} J. K. Stockton, K. Takase, and M. A. Kasevich, Phys. Rev. Lett. 107, 133001 (2011).
\bibitem{Hallwood2006a} D. Hallwood, K. Burnett, and J. Dunningham, New J. Phys. {\bf 8} , 180 (2006).
\bibitem{Giovannetti2006a} V. Giovannetti, S. Lloyd, and L. Maccone, Phys. Rev. Lett. {\bf 96}, 010401 (2006). 
\bibitem{Giovannetti2004} V. Giovannetti, S. Lloyd and L. Maccone, Science \textbf{306} 1330 (2004).
\bibitem{Dowling2008} J.P. Dowling {Contemp. Phys.} {\bf 49}, 125 (2008).
\bibitem{Pezze2009a}  L. Pezz\'e and A. Smerzi, {Phys. Rev. Lett.} {\bf 102}, 100401 (2009).
\bibitem{Holland1993a} M.J. Holland and K. Burnett, {Phys. Rev. Lett.} {\bf 71}, 1355 (1993).
\bibitem{Dunningham2002a} J.A. Dunningham, K. Burnett and S.M. Barnett, Phys. Rev. Lett. {\bf 89}, 150401 (2002).
\bibitem{Gerry} R.A. Campos, C.C. Gerry and A. Benmoussa, {Phys. Rev.} A \textbf{68}, 023810 (2003).
\bibitem{Giovannetti2011} V. Giovannetti, S. Lloyd and L. Maccone, {Nature Photon.} {\bf 5}, 222 (2011).
\bibitem{Higgins2007} B.L. Higgins, D.W. Berry, S.D. Bartlett, H.M. Wiseman and G.J. Pryde {Nature} \textbf{450}, 393 (2007).
\bibitem{Kacprowicz2010a} M. Kacprowicz, R. Demkowicz-Dobrzanski, W. Wasilewski, K. Banaszek and I.A. Walmsley, Nature Photon. {\bf 4}, 357 (2010).
\bibitem{Oberthaler} J. Esteve, C. Gross, A. Weller, S. Giovanazzi and M.K. Oberthaler, {Nature} {\bf 455}, 1216 (2008).
\bibitem{Cooper2012a} J.J. Cooper, D.W. Hallwood, J.A. Dunningham, and J. Brand, Phys. Rev. Lett. {\bf 108}, 130402 (2012).



\bibitem{Wolfgramm2013a} F. Wolfgramm et al., Nature Photon. {\bf 7}, 28 (2013).
\bibitem{Aasi2013a} J. Aasi {\it et al.}, Nature Photon. {\bf 7}, 613 (2013). 
\bibitem{feder} A.G. Morris and D.L. Feder, Phys. Rev. A {\bf 74}, 033605 (2006).

\bibitem{dagnino} D. Dagnino, N. Barber\'an, M. Lewenstein and J. Dalibard, Nature Phys. \textbf{5}, 431 (2009).

\bibitem{ricoNJP} L.M. Rico-Gutierrez, T.P. Spiller and J.A. Dunningham, {New. J. Phys} \textbf{15} 063010 (2013).
\bibitem{Cooper2010a} J.J. Cooper, D.W. Hallwood and J.A. Dunningham, Phys. Rev. A {\bf 81}, 043624 (2010).
\bibitem{Caves} S.L. Braunstein and C.M. Caves, {Phys. Rev. Lett.} {\bf 72}, 3439 (1994).
\bibitem{Feshbach} W.C. Stwalley, Phys. Rev. Lett. {\bf 37}, 0031 (1976).
\bibitem{messiah} Albert Messiah {\it Quantum Mechanics Volume II}, 750--753 (1965).
\bibitem{soding} J. Soding, D. Guery-Odelin, P. Desbiolles, F. Chevy, H. Inamori and J. Dalibard, {Appl. Phys. B} {\bf 69}, 257 (1999).
\bibitem{gemelke} N. Gemelke, E. Sarajlic and S. Chu, arXiv:1007.2677 (2010). 

\bibitem{Fazio} A. Osterloh, L. Amico, G. Falci, and R. Fazio, Nature Phys. \textbf{416} 608 (2002).
\bibitem{Zanardi} P. Zanardi, P. Giorda, and M. Cozzini, Phys. Rev. Lett. {\bf 99}, 100603 (2007).
\bibitem{ZanardiTwo} P. Zanardi, M.G.A. Paris, and L. C. Venuti, Phys. Rev. A {\bf 78}, 042105 (2008).
\bibitem{HollandBurnett} M.J. Holland and K. Burnett, Phys. Rev. Lett. {\bf 71}, 1355 (1993).
\bibitem{DalibardAngMom} F. Chevy, K.W. Madison and J. Dalibard, Phys. Rev. Lett. {\bf 85}, 2223 (2000).
\bibitem{GemelkeChu} E. Sarajlic, N. Gemelke, S.W. Chiow, S. Herrman, H. M\"uller and S. Chu {\it Pushing the Frontiers of Atomic Physics}, 34--45 (2009).
\bibitem{Fetter2009} A.L. Fetter, Rev.Mod.Phys. {\bf 81}, 647 (2009).
\bibitem{DemkowiczReview} R Demkowicz-Dobrzanski, M. Jarzyna and J. Kolodynski, arXiv:1405.7703(2014). 


%
%
%
%








%






%


%

\end{thebibliography}
\end{document}